\documentclass[12pt]{amsart}
\usepackage{times}
\usepackage{amssymb, amsmath}
\usepackage{amsthm}
\usepackage{tikz}
\usepackage{bm}
\usetikzlibrary{matrix,arrows}
\usepackage{enumitem}
\usepackage{booktabs}
\usepackage[section]{placeins}
\usepackage{subcaption}

\theoremstyle{plain}

\theoremstyle{remark}

\graphicspath{{./images/}}

\usepackage[a4paper, total={6in, 8in}]{geometry}

\begin{document}

\title[Feature selection for intrusion detection systems]{Feature selection for intrusion detection systems}
\author[F. Kamalov, S. Moussa, R. Zgheib, O. Mashaal]{Firuz Kamalov$^1$$^{\boldsymbol{*}}$, Sherif  Moussa$^2$, Rita Zgheib$^3$, Omar Mashaal$^4$}

\address{$^{1}$ Canadian University Dubai, Dubai, UAE.}
\email{\textcolor[rgb]{0.00,0.00,0.84}{firuz@cud.ac.ae}}

\address{$^{2}$ Canadian University Dubai, Dubai, UAE.}
\email{\textcolor[rgb]{0.00,0.00,0.84}{smoussa@cud.ac.ae}}

\address{$^{3}$ Canadian University Dubai, Dubai, UAE.}
\email{\textcolor[rgb]{0.00,0.00,0.84}{rita.zgheib@cud.ac.ae}}

\address{$^{4}$ Canadian University Dubai, Dubai, UAE.}
\email{\textcolor[rgb]{0.00,0.00,0.84}{omar.mashaal@cud.ac.ae}}

\date{Nov 11, 2020
\newline \indent $^{\boldsymbol{*}}$ Corresponding author
\newline \indent DOI: 10.1109/ISCID51228.2020.00065}

\begin{abstract}
In this paper, we analyze existing feature selection methods to identify the key elements of network traffic data that allow intrusion detection. In addition, we propose a new feature selection method that addresses the challenge of considering continuous input features and discrete target values. We show that the proposed method performs well against the benchmark selection methods. We use our findings to develop a highly effective machine learning-based detection systems that achieves 99.9\% accuracy in distinguishing between DDoS and benign signals. We believe that our results can be useful to experts who are interested in designing and building automated intrusion detection systems.
\end{abstract}

\maketitle

\section{Introduction}
Network-based technologies have permeated every aspect of our daily lives. However, the great benefits of network-based technologies come at a cost of increased vulnerability to malicious attacks. The consequences of a malicious attack on a network have become as devastating as ever. A well executed  attack on a power utility's network can leave millions of people without electricity. As a result it has become vital to develop effective intrusion detection systems (IDS) that can ensure the safety of a network. Recent advances in machine learning algorithms have made it possible to develop IDS using artificial intelligence. Machine learning algorithms have the capability to independently build detection systems using only the raw traffic data. A number of studies have already shown the effectiveness of machine learning in detecting malicious attacks on a network. One of the key aspect of developing a detection system using machine learning is selection of  appropriate features of the data to use for model building. Given a dataset with a large number of features, identifying the right features for model building can dramatically improve the outcome. 

Machine learning is increasingly being used to automate many IDS processes that were previously done by hand including discovering and adding suspicious URL to the blocked list, adding new rule, creating connection exceptions, an others. If done manually, such activities represent a huge burden on the system administrator. One of the main advantages of machine learning is its scalability. Continued growth of network size and traffic together with increasing number of zero-day attacks and other online threats have made automated machine learning IDS an indispensable tool in defending against malicious attacks. ML-based IDS has the capability to learn from pervious intrusion attempts to identify the malicious traffic patterns associated with them, any future occurrence of these attacks will be identified and classified much faster than human based systems regardless of the size of the network.

In this paper, our goal is to study feature selection in network traffic data with the aim of detecting potential attacks. We consider various existing feature selection methods as well as propose a new feature selection algorithm to identify the most potent features in network traffic data. Our results have a dual application. First, we shed light on the key features in network traffic data that help detect a malicious signal. The properties of malicious signals identified by the feature selection methods can help IT professionals to better protect their networks. Second, feature selection is used to develop effective machine learning-based IDS. We present several detection algorithms built with the help of feature selection methods that achieve remarkable performance with respect to detection and false alarm rates.

To address the difference between continuous and discrete features, we propose a novel forward search algorithm that combines correlation and mutual information to select the optimal features. Concretely, we use the linear correlation coefficient to measure the redundancy between a new feature and the existing feature subset. And we use mutual information to measure the relevance of the new feature to the target variable. The final feature score is calculated based on a weighted difference between the relevance and redundancy of the feature. 

Our paper is structured as follows. In Section 2, we present an overview of existing literature related to intrusion detection and machine learning. 
In Section 3, we provide the description of the proposed algorithm.
In Section 4, we carry out numerical analysis of the existing and the proposed feature selection techniques in the context of network security. 
Section 5 concludes the paper with a brief summary and discussion of future work.
\section{Literature review}

IDS use one of three detection methods to identify malicious activities, signature-based detection compare the possible attack signature with stored signature of previous attacks \cite{Santoso}. This method has a low rate of false positive and is very effective. However, if the attack signature does not exist in the signature database it is less likely to be detected. In other words, signature-based detection is not efficient against zero-day attacks. The second detection method is anomaly detection in which any abnormal behavior that deviates from regular baseline operation is detected \cite{Tama}. This method outperforms the signature-based approach in detecting zero-day attacks due to the uniqueness of the operational baseline for individual networks. Despite their advantages anomaly detection methods suffer from the high false positive alarms due to the fact that some of the detected baseline deviations are legitimate network behavior. The third detection method is the hybrid detection. It is a combination of the above two methods used to alleviate the weaknesses of the signature based and anomaly techniques. To reduce zero-day attacks and high false positive problems, multiple algorithms must be processed concurrently to decide about the anomality of an event and simultaneously match the signature to the previously recorded attack \cite{Garg}.

Machine learning methods are classified into two types based on the training method. First, supervised learning algorithms are designed to be trained on labeled data. There exists a variety of supervised algorithms that have been used in IDS including Logistic Regression \cite{Shah}, Gaussian Naive Bayes \cite{Singh}, Support Vector Machine (SVM), Random Forest \cite{Ahmad} and others. The second type of ML algorithms are the unsupervised learning algorithms which do not rely on labeled data for training and classification. Unsupervised algorithms have the ability to discover patterns and classify traffic automatically. Unsupervised algorithms are particularly useful when labeled training data is unavailable. There exists a variety of supervised algorithms that have been used in IDS including K-Means Clustering, Single Linkage Clustering \cite{Su}, Quarter-sphere Support Vector Machine \cite{Rajasegarar} and others. 

Another important aspect of ML-based IDS is the ability to process large amount of data quickly in order to detect attacks in real time, classification process consumes large amount of computational time due to the high number of features that must be classified. In fact, not all features are useful to the classification process and some features are either considered as noisy and degrade the process performance or highly correlated to each other and could be omitted \cite{Zuech}. Feature selection techniques are used to improve the classification performance, many researchers tackled the feature selection algorithms in order to minimize the number of selected features by choosing only the most significant features \cite{Ambusaidi}. As a result, ML-based IDS can perform predictions more efficiently. 

Feature selection is an important preprocessing step in many machine learning tasks. One of the most popular selection techniques is filter approach. It is a simple and efficient method that has been shown to perform effectively in a number of situations. In the filter approach, a univariate measure of feature importance is used to rank and select features. In case of continuous features, commonly used metrics include correlation and $F$-statistics. For discrete features, researchers use $\chi^2$-statistic and mutual information. In general, there exists a number of option for feature evaluation metrics. As shown in \cite{Kamalov4}, any measure that satisfies generalized similarity criteria can be employed in feature evaluation.

In \cite{Amiri}, the authors consider the issue of feature selection in the context of IDS. They propose two separate forward search selection algorithms based on linear correlation coefficient and mutual information. Our method differs crucially in that we propose to combine the two metrics into one.
Recently, more advanced approaches to basic univariate selection method have been developed. In \cite{Kamalov1}, the authors combine mutual information and $\chi^2$-statistic into a vectorized feature score. In \cite{Kamalov2}, the authors employ Sobolev variance decomposition to evaluate features. The proposed method allows to account for feature interactions under certain conditions. The authors in \cite{Kamalov3} reveal monotonicity property of $\chi^2$ statistic and use it to improve the feature selection process. They show that a new feature can be added to an existing feature subset only if the $\chi^2$ values increases beyond a certain threshold. 
\section{MICorr selection method}
One of the main challenges in filter approach is dealing with different types of variables. In particular, there is no consensus in the literature on how to deal with continuous input and discrete output variables. Relationship between continuous variables is usually measured based on correlation. On the other hand, discrete variables employ entropy based metrics. Existing feature selection method use a uniform approach to all variable ignoring the data type. For instance, continuous variables are binned and treated as discrete.

The problem of mixed variables arises prominently in the context of intrusion detection. Most of the variables inside  network traffic data are continuous (flow duration, number of packets, length of packets, etc) whereas the target variable is discrete (benign/malicious). Therefore, an effective solution of the problem has significant implications in intrusion detection.
We propose a new method (MICorr) to address the above issue by combining the Pearson correlation and mutual information. 
Given a feature variable $x$ and target variable $y$, we calculate the feature score by taking the difference between the feature relevance and feature redundancy

\begin{equation}
\mbox{MICorr}(x) = \mbox{Rel}(x, y) - \mbox{Red}(x, S),
\end{equation}
where $S$ is the subset of already selected features. The MICorr score attempts to simultaneously maximize the information shared between the feature and the target variable while minimizing the correlation with the already selected features. As a results the selected feature will contribute the maximum amount of new information to the optimal feature subset.
Let $F$ denote the full set of features. The algorithm for executing the proposed selection method is provided below:
\\
\\
\textbf{Algorithm}
\\
\line(1,0){150}
\begin{enumerate}
\item Initialize $S=\emptyset$
\item Choose $x_0$ with max Rel and add it to $S$. Remove feature $x_0$ from $F$.
\item For each $x$ in $F$, calculate MICorr($x$) and select the feature $x_1$ with the highest score. Add $x_1$ to $S$ and remove $x_1$ from $F$.
\item Repeat Step 2 until the desired number of features is selected.
\end{enumerate}
\section{Results and analysis}
Our goal in this section is to analyze various feature selection methods in the context of intrusion detection. In addition, we demonstrate the effectiveness of the newly proposed feature selection method in identifying the characteristics of network traffic data that are relevant in intrusion detection. We run our numerical experiments on the CSE-CIC-IDS2018 on AWS dataset that consists of 10,000 benign and DDoS instances. Experiments reveal that feature selection helps build potent IDS using machine learning techniques.
\subsection{Data}
The dataset used in our experiment is obtained from the collaborative project between the Communications Security Establishment and the Canadian Institute for Cybersecurity \cite{Sharafaldin}. The data is based on the creation of user profiles which contain abstract representations of events and behaviors seen on the network. It is based on a simulated LAN network topology that is common on the AWS computing platform.  We extract a random sample of size 10,000 from the original data. The data is evenly distributed between benign and DDoS instances. Each instance of the dataset consists of 63 continuous features and a label. As shown in Table \ref{TP}, the features consists of various packet and IAT characteristics.  Our goal is to apply machine learning techniques to identify the most relevant features for intrusion detection.

\subsection{Feature selection methods}
In our study of relevant features we employ three standard selection methods: correlation-based univariate, MI-based univariate, and correlation-based forward search algorithms. Correlation (MI) based univariate methods simply rank features based on their correlation (MI) with the target variable. Then the desired number of features is selected based on the ranking. Forward search correlation-based method iteratively selects features based on maximum relevance and minimum redundancy. The relevance is calculated based on the correlation between the feature and the target variable while redundancy is calculated based on the correlation between the feature and the previously selected subset of features. The above algorithms are quick tools to process the data prior to classification. In addition, we apply the MICorr selection method to the IDS data.
\subsection{Results}
The feature rankings based on various selection criteria are presented in Table \ref{tab:top}. The columns U-MI and U-Corr represent the univariate selection methods and FS-Corr represents the forward search correlation-based method. Note that MICorr and U-MI share more than half of the selected features. On other hand, FS-Corr and U-Corr have only a few features in common. We also note that feature 6 (Fwd Packet Length Max) and feature 44 (Average Packet Size) are shared by three selection methods indicating their potential significance in IDS. The top two features shared by MICorr and U-MI are 44 and 49 (Subflow Fwd Bytes). 

The analysis of the results in Table \ref{tab:top}, reveal that features 6, 44, and 49 are the most important in helping detect DDoS attacks on a network. 
In networks, flow is a series of packets between a source IP and port to a destination IP and port. Forward flow is when a packet moves from node A to node B, and backward is when the reply for the packet moves from node B to node A. 
It is essential to notice that the Packet Length specifies the whole packet's size, including the header, trailer, and data. However, the Packet Size determines only the size of the header on the packet. The results show that the difference of the packet size and length between the attack packets and the legitimate packets can be used to distinguish attack traffic from legitimate traffic. Indeed, attackers usually generate small packets or even empty packets to reduce the resources required. However, the packets of communication protocols are also small, packets filled with user data, which usually are large, make up a considerable percentage of the legitimate traffic. 

\begin{table}[h!]
\centering
\caption{The top 10 features selected by each features selection method.}
\label{tab:top}
\begin{tabular}{lrrrrr}
\toprule
{Rank} &  MICorr & U-MI  &  FS-Corr &  U-Corr &  Tree importance \\
\midrule
1  &    44 &        44 &      13 &          13 &             45 \\
2  &    49 &        49 &      41 &          36 &              6 \\
3  &    18 &         4 &      55 &          12 &             37 \\
4  &     4 &         8 &      23 &          46 &             36 \\
5  &     5 &         6 &      12 &          10 &              8 \\
6  &    52 &        45 &      11 &          34 &              4 \\
7  &    21 &        18 &       6 &          44 &             10 \\
8  &     6 &        35 &      36 &          37 &             54 \\
9  &    35 &        21 &       7 &          35 &             13 \\
10 &     8 &        34 &      32 &          55 &             20 \\
\bottomrule
\end{tabular}
\end{table}

 
Machine learning has been widely used in a number of applications including design of detection systems. Although there exists a number of powerful ML techniques including Deep Learning and SVM they are black box models that do not reveal the internal decision making process of the model. In contrast, Decision Trees and Random Forest offer competitive alternatives that allow users analyze the internal structure of the model. Since understanding the decision making process of an algorithm is critical in security related environment, we employ Random Forest as the ML model in our experiments. Random Forest is an ML algorithm based on a collection of Decision Trees. It is trained on multiple subsets of the original data and the final prediction is made based on a voting criterion among all weak classifiers. As a result, we obtain a classifier with low bias and variance.

\begin{figure}[h!]
\begin{subfigure}{1\textwidth}
  \centering
  \includegraphics[width=1\linewidth]{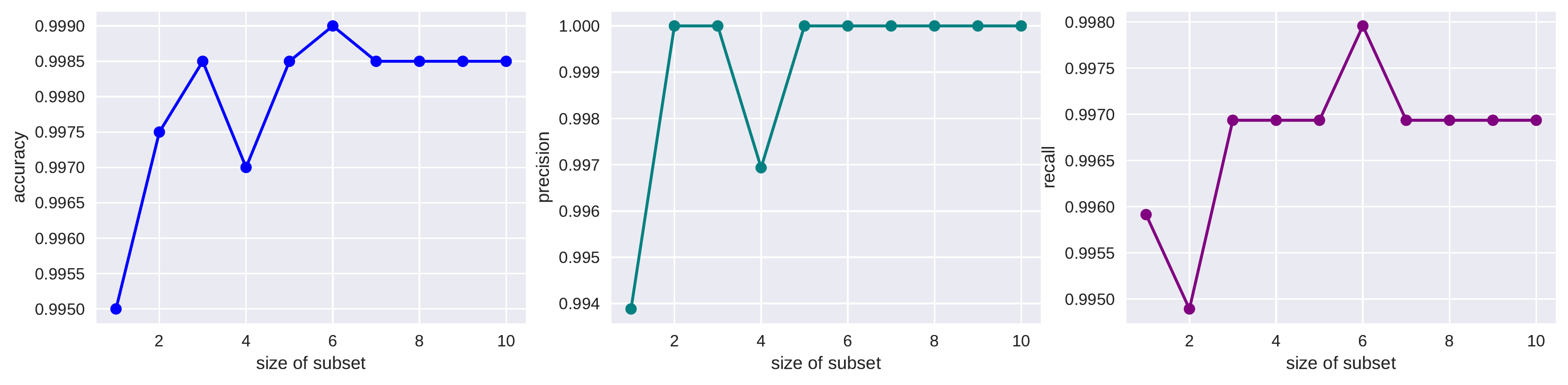}  
  \caption{Performance results using MICorr feature selection.  }
  \label{fig:forward_MI}
\end{subfigure}
\newline
\begin{subfigure}{1\textwidth}
  \centering
  \includegraphics[width=1\linewidth]{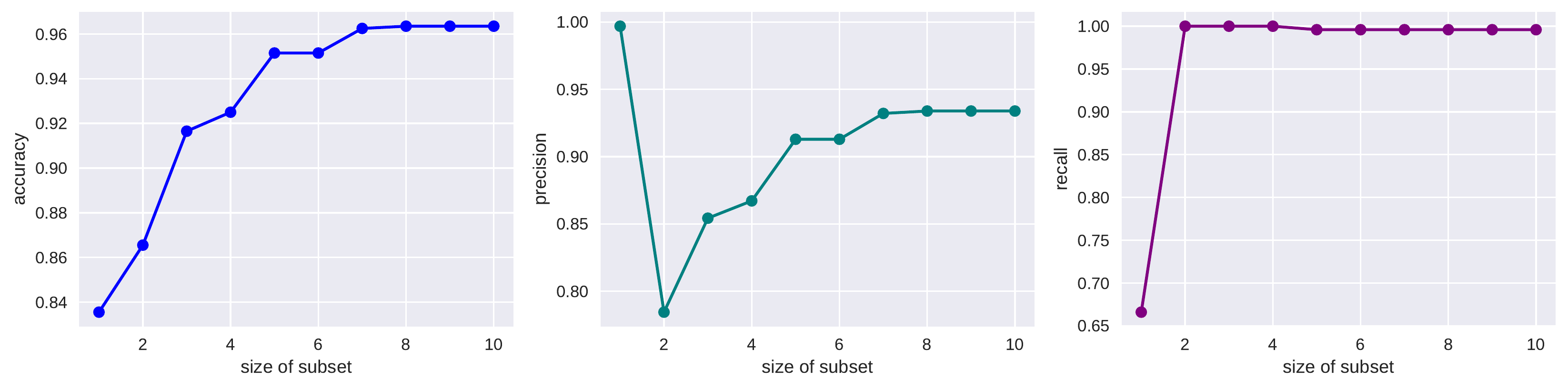}  
  \caption{Performance results using FS-Corr feature selection. The accuracy of the classifier increases with the size of the selected feature subset.}
  \label{fig:forward_correlation}
\end{subfigure}
\newline
\begin{subfigure}{1\textwidth}
  \centering
  \includegraphics[width=1\linewidth]{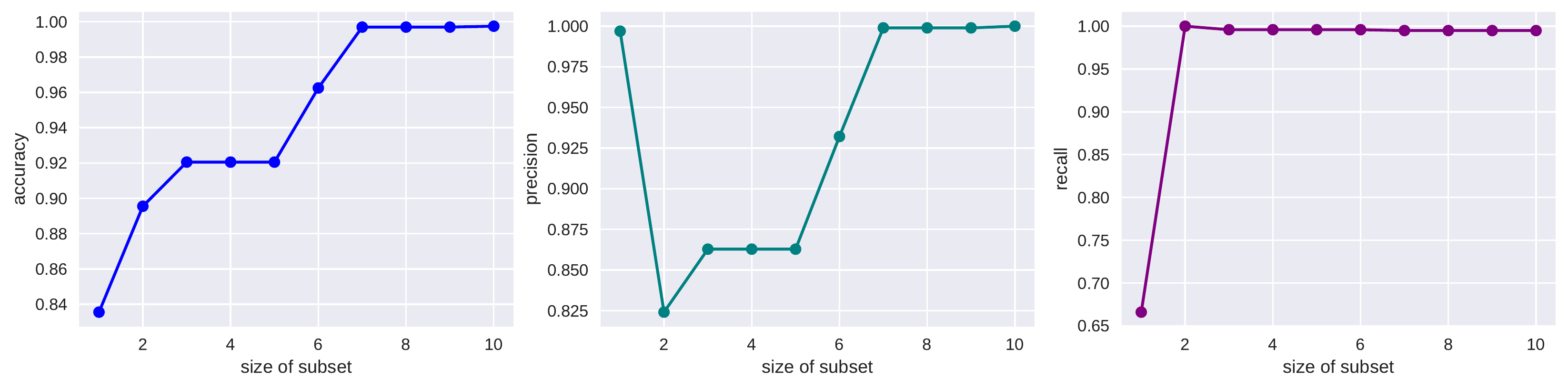}  
  \caption{Performance results using U-Corr feature selection. }
  \label{fig:univariate_correlation}
\end{subfigure}
\newline
\begin{subfigure}{1\textwidth}
  \centering
  \includegraphics[width=1\linewidth]{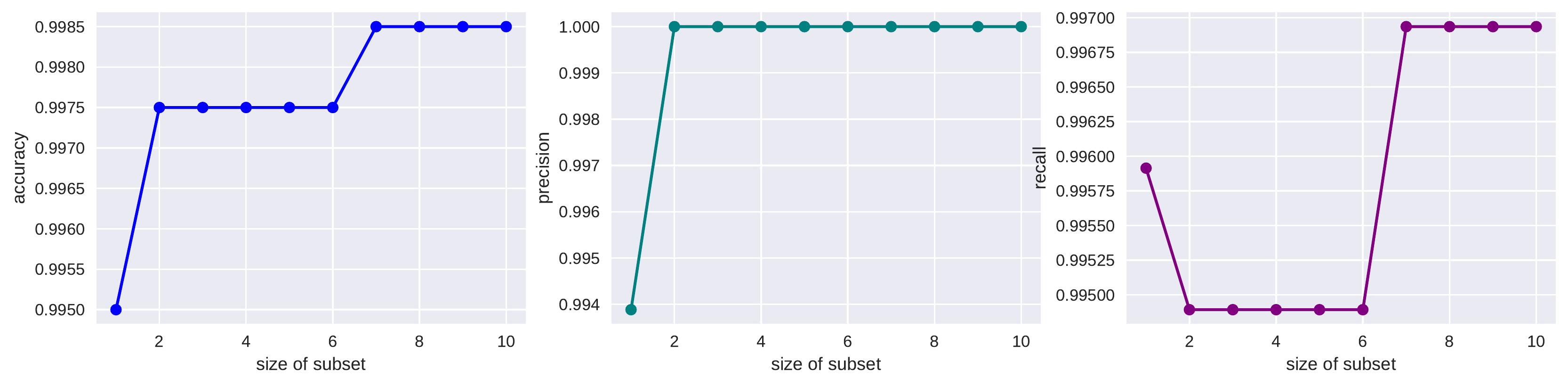}  
  \caption{Performance results using U-MI feature selection.}
  \label{fig:univariate_MI}
\end{subfigure}

\caption{Performance comparison of various feature selection methods.}
\label{fig:results}
\end{figure}

For each feature selection algorithm,  we used different subsets of selected features to train a RF model. The subsets were trained and tested on the IDS2018 dataset according to 80/20 split. The performance of each subset was measured using model accuracy, precision, and recall. The results of the experiment are presented in Figure \ref{fig:results}. As shown in Figure \ref{fig:forward_MI}, an IDS model built with the top 6 MICorr-selected features is capable of reaching $0.999$ accuracy rate. Furthermore, he top 6 MICorr-selected features produce a $1.00$ precision and $0.998$ recall rates.
The results show that the proposed selection method is capable of identifying relevant features in network traffic data and train a powerful ML-based IDS. Other feature selection methods tested in the experiment produce similarly good results albeit slightly higher number of features than MICorr. We note that the top 7 U-Corr selected features achieve $1.00$ accuracy rate, but the top 6 features achieve only $0.96$ accuracy.

\begin{figure}
\centering
\includegraphics[width=1\textwidth]{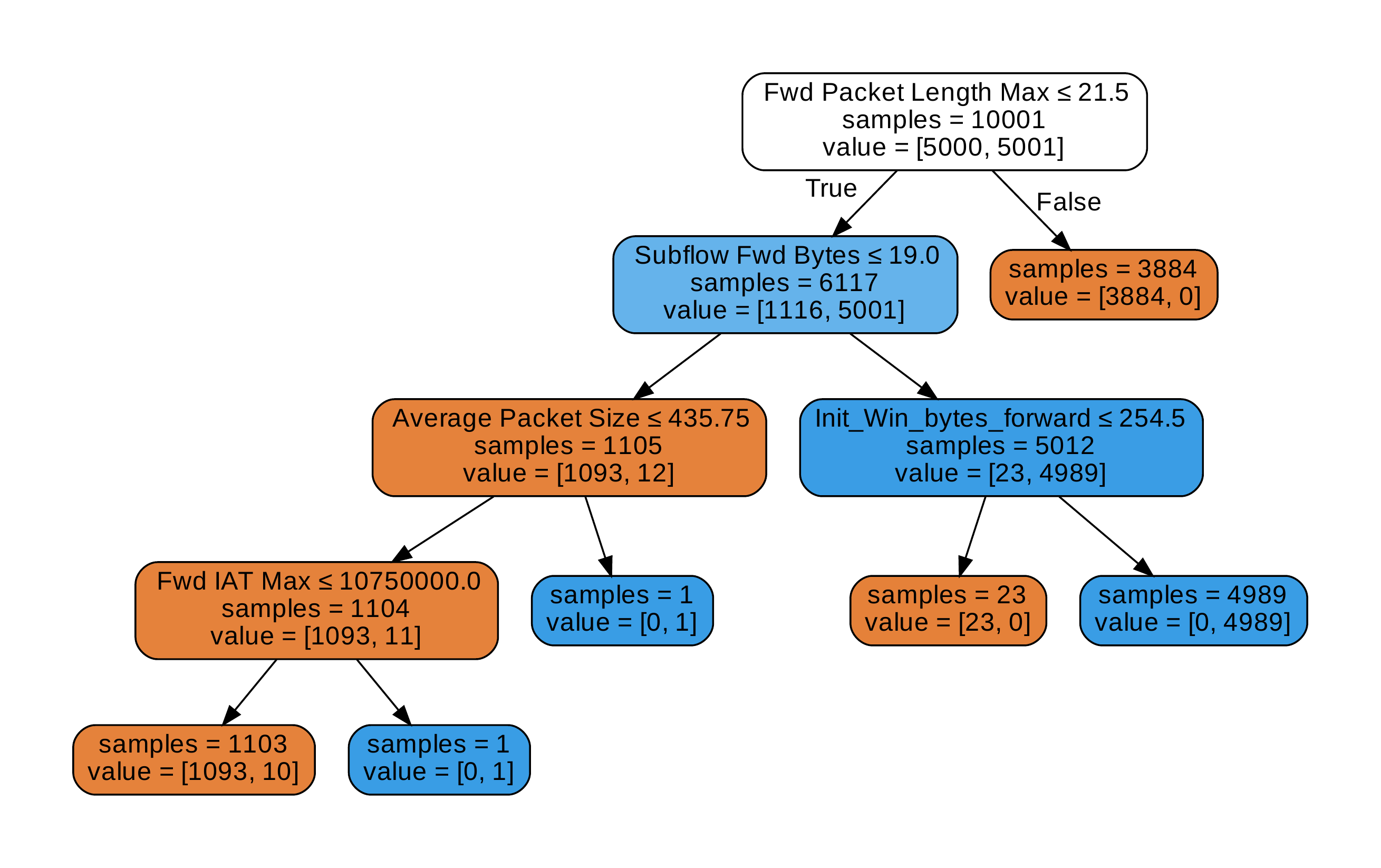}
\caption{Decision Tree based on the MICorr-selected features. Blue and orange nodes indicate DDoS and benign instances respectively. The model achieves 99.9\% classification accuracy.}
\label{mi_tree}
\end{figure}

We apply the Decision Tree algorithm to construct an IDS model based on the features selected with MICorr method. As shown in Figure \ref{mi_tree}, the model accurately classifies 99.9\% of all instances. We can also see that the most relevant features in the tree are Fwd Packet Length Max, Subflow Fwd Bytes, Average Packet size, and Fwd IAT Max. 
Besides the importance of the length and size of packets and bytes in distinguishing benign from malignant packets, the Decision Tree algorithm with MICorr shows that the time (IAT Max) is also essential. Time-dependent sequential data retains the knowledge of the previous packet’s effect on the current packet. For a particular time $t$, continuous network packets are captured to form an input window. An algorithm exhibited in legitimate and malignant packets from some historical data can differentiate between the benign and malignant incoming packets \cite{Diro}.

\section{Conclusion}
The advent of ubiquitous network based technologies has increased the associated vulnerabilities. As a result, it has become paramount to design and implement effective IDS. In this paper, we apply feature selection methods to improve our understanding of relevant features inside network traffic data and construct potent detection systems. We examine three standard filter selection methods as well as introduce a new selection method.
The proposed selection method aims to addresses the discrepancy between continuous input features (Packet Length, Subflow Bytes, etc) and discrete target variable (benign/malicious). The results of numerical experiments show that our method achieve a high degree of accuracy (99.9\%) in distinguishing between benign and malicious signals. 

We believe that feature selection can improve our understanding of traffic data and its characteristics in the context of intrusion detection. It also helps to design better IDS using ML techniques. In future, it would be important to expand the scope of our study to other types of attacks.

\begin{table}
\centering
\caption{The table of all the features in the dataset.}
\label{TP}
\small
\begin{tabular}{llll}
\toprule
Index &                        Name &  Index &                   Name  \\
\midrule
1  &                 Flow Duration &        33 &        Min Packet Len  \\
2  &             Total Fwd Packets &       34 &        Max Packet Len    \\
3  &        Total Bwd Packets &       35 &       Packet Len Mean  \\
4  &   Total Len Fwd Packets &      36 &        Packet Len Std \\
5  &   Total Len of Bwd Packets &        37 &   Packet Len Variance \\
6  &         Fwd Packet Len Max &       38 &           FIN Flag Count   \\
7  &         Fwd Packet Len Min &       39 &           SYN Flag Count     \\
8  &        Fwd Packet Len Mean &       40 &           PSH Flag Count        \\
9  &         Fwd Packet Len Std &        41 &           ACK Flag Count     \\
10 &         Bwd Packet Len Max &      42 &           URG Flag Count        \\
11 &         Bwd Packet Len Min &       43 &             Down/Up Ratio      \\
12 &        Bwd Packet Len Mean &        44 &       Average Packet Size         \\
13 &         Bwd Packet Len Std &       45 &      Avg Fwd Seg Size       \\
14 &                 Flow IAT Mean &       46 &      Avg Bwd Seg Size     \\
15 &                  Flow IAT Std &       47 &       Fwd Header Length     \\
16 &                  Flow IAT Max &       48 &       Subflow Fwd Packets     \\
17 &                  Flow IAT Min &          49 &         Subflow Fwd Bytes     \\
18 &                 Fwd IAT Total &       50 &       Subflow Bwd Packets    \\
19 &                  Fwd IAT Mean &     51 &         Subflow Bwd Bytes      \\
20 &                   Fwd IAT Std &       52 &    Init\_Win\_bytes\_fwdd     \\
21 &                   Fwd IAT Max &       53 &   Init\_Win\_bytes\_bwd     \\
22 &              Fwd IAT Min &			54 &          act\_data\_pkt\_fwd  \\
23 &            Bwd IAT Total & 			55 &      min\_seg\_size\_fwd 	\\
24 &             Bwd IAT Mean &		56 &               Active Mean	\\
25 &              Bwd IAT Std & 			 57 &                Active Std	\\
26 &              Bwd IAT Max &  		 58 &                Active Max	\\
27 &              Bwd IAT Min & 			59 &                Active Min		\\
28 &            Fwd PSH Flags &		 60 &                 Idle Mean 	\\
29 &        Fwd Header Len &		61 &                  Idle Std 	\\
30 &        Bwd Header Len &		62 &                  Idle Max 	\\
31 &            Fwd Packets/s & 		  63 &                  Idle Min	\\
32 &            Bwd Packets/s & {} &{}\\

\bottomrule
\end{tabular}
\end{table}

\end{document}